# Rapid Control Selection through Hill-Climbing Methods

Krispin A. Davies, Alejandro Ramirez-Serrano, Graeme N. Wilson, Mahmoud Mustafa

Dept. of Mechanical and Manufacturing Engineering, University of Calgary, 2500 University Drive, Calgary, Canada

krispin.davies@live.ca, {aramirez, wilsongn, mmmustaf}@ucalgary.ca

**Abstract.** Consider the problem of control selection in complex dynamical and environmental scenarios where model predictive control (MPC) proves particularly effective. As the performance of MPC is highly dependent on the efficiency of its incorporated search algorithm, this work examined hill climbing as an alternative to traditional systematic or random search algorithms. The relative performance of a candidate hill climbing algorithm was compared to representative systematic and random algorithms in a set of systematic tests and in a real-world control scenario. These tests indicated that hill climbing can provide significantly improved search efficiency when the control space has a large number of dimensions or divisions along each dimension. Furthermore, this demonstrated that there was little increase in search times associated with a significant increase in the number of control configurations considered.

## 1    Introduction

The past decade has witnessed a migration of autonomous systems (robots) from controlled environments, such as laboratories and factories, to the uncontrolled environments of everyday life, such as mining and other exploration areas. Diverse devices such as autonomous cars and the recently announced DARPA challenge focusing on humanoid search and rescue robots are now under development, [1,2]. These everyday scenarios present three main challenges over their controlled counterparts: i) transient vehicle models, ii) uncontrolled environments, and iii) multiple goals. The transient vehicle model refers to the vehicle dynamics and limitations (both physical limits and control bounds) which may vary in time, [3]. The second challenge, the uncontrolled environment, presents both a wide range of terrain properties (e.g., surface stiffness and frictional coefficient) and obstacles (both static and dynamic), [4]. Finally, multiple goals, potentially competing or even mutually exclusive, may exist, [5]. Traditional direct control techniques (e.g. adaptive control and optimal control) become highly complex when faced with these challenges, because they use algebraic equations to map a vehicle state, $\vec{s}$, and set of goals, $g(\vec{s})$, directly to a control configuration, $\vec{c}$, (i.e., a set of control signals completely defining the vehicle action), [6]. It is difficult to create equations that account for all possible combinations of input factors these challenges can produce. As an alternative, the control problem can be framed as an optimization problem within the space of all possible control configurations, $\vec{c}_i \in \vec{\mathcal{C}}$. Within the control space, a search algorithm attempts to identify the best control configuration, $\vec{c}_{best}$, given the current goals and constraints as evaluated in a heuristic (e.g., cost) function, $h(\vec{s}, \vec{c}_i)$. Herein produces a form of model predictive control (MPC) as employed in process industries and more recently in robotics, [7].

Employing MPC creates a three part loop structure in place of a direct control law (Fig. 1) where a search algorithm suggests a control configuration, $\vec{c}_i$. The suggested control configuration is then used by a simulation algorithm to predict how the vehicle will behave (e.g., move) over a predetermined interval. The heuristic algorithm compares the predicted vehicle behavior to the defined goals in order to produce a single numerical cost, $h(\vec{s}, \vec{c}_i)$. Finally the cost, is used as feedback by the search algorithm when selecting the next control configuration, $c_{i+1}$.

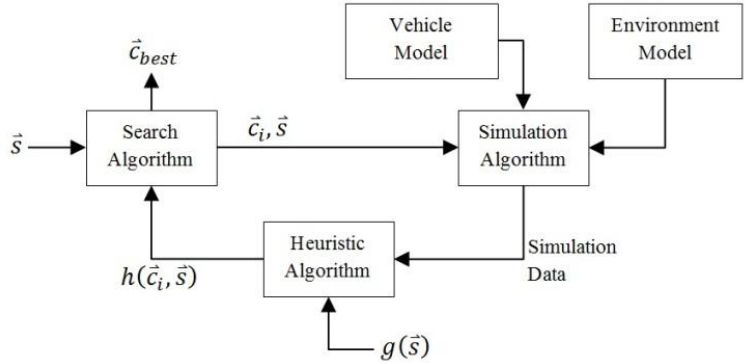

**Fig. 1.** Model Predictive Control Cycle

This sidesteps the need for direct control equations, and provides greater freedom. First, as the control configuration is interpreted a variety of formats can be used. For example, as a time polynomial for each control surface angle on an aerial vehicle or as a set of turning angles to be executed in sequence by an autonomous car, [3,8]. Likewise, the goals (e.g., user defined) are also interpreted granting freedom in their representation, for example: as linearly increasing time cost and as a step cost for potential collisions, [9]. Finally, the vehicle and environment models are only used in the simulation algorithm for which efficient techniques already exist, [10]. Thus, framing the control problem as MPC provides significant benefits at the expense of the time required to execute the search.

The time cost associated with simulation and heuristic evaluation is the main drawback of the MPC approach. Based on work conducted in this area, a search-simulate-evaluate loop cycle may take longer to process than an entire direct control calculation, [11]. Thus, minimizing the number of search cycles needed to select a suitable control configuration is crucial. In this case, a suitable control configuration could mean the globally optimal control configuration or simply a control configuration equal to or better than a predetermined benchmark. In an effort to reduce this search time, we propose the use of hill climbing search techniques instead of the systematic or random search algorithms employed throughout literature. It is envisioned that hill-climbing, being a subset of gradient decent, will improve performance as many of the heuristic functions produced by MPC are largely continuous, [12].

Thus, in order to determine whether the hill climbing group of search algorithms is suitable for MPC, this work compared random restart hill climbing's (RRHC) performance to a representative systematic search algorithm (progressive grid refinement search) and a representative random search algorithm (pure random search). As RRHC can be implemented readily and with computational overhead comparable to systematic and random search algorithms it is suitable for comparison, [13]. All three search algorithms are described briefly in Section 2. Initial testing of the search algorithms was conducted using a systematically defined control space heuristic, described in Section 3, to examine how performance is affected by key search parameters. As the systematic testing produced unexpectedly positive results, Section 4 applies these search algorithms to a real-world vehicle control selection problem in order to validate the systematic testing results. Finally, we present our conclusions on the potentials of both MPC in robotics and the use of hill climbing algorithms within MPC (Section 5).

## 2    Search Methodologies

The purpose of the search algorithm is to identify a suitable control configuration as rapidly as possible. In most scenarios, "suitable" may encompass several control configurations, all of which meet or exceed a bench mark rating. As heuristics are often evaluated on a basis of cost, in this paper it is assumed that lower heuristic costs are better. Thus, for any given control problem, there is a set of one or more control configurations which are suitable. However the number and location(s) of this/these solution(s) is/are unknown prior to conducting the search. This realization was used to select the three search algorithms: RRHC, Grid Refinement, and Random Sampling.

### 2.1    Hill Climbing Algorithm

Hill climbing methods were selected for their ability to rapidly identify minima within continuous spaces. As heuristic functions can be constructed to be generally continuous, though they may contain micro-discontinuities, gradient methods are suitable for solving these functions. Given the inability to analytically determine the heuristic's gradient due to the dependence on simulation results however, a numerical approach such as hill climbing is needed. Of the hill climbing methods, RRHC most closely resembled the literature methods in computational overhead making it most suitable for comparison.

Functionally, the core of RRNNHC lies in the nearest climb behavior. In this, the current control configuration's cost, $h_{current}$, is compared with the cost of those configurations closest to it within the control space. This comparison has two possible results: i) neighboring cost < current cost, or ii) neighboring cost ≥ current cost. If any of the neighboring configurations has a lower cost than the current configuration, then the current configuration is replaced by the neighboring configuration with the lowest cost. This produces the gradient descent behavior. Conversely, if the current configuration has a lower cost than all neighboring configurations, then it represents a minimum. Upon reaching a minimum, random restart is replaces the current control configuration with a randomly selected control configuration from which gradient descent can begin again.

### 2.2    Comparison Algorithms

The hill climbing algorithm's performance was compared with two competing algorithms (systematic searches and random based searches), each representing a method common throughout the literature. The first of these algorithms was grid refinement, which systematically searched the entire control space according to a series of progressively finer and finer grids, each with $2^i$ grid divisions where $i = [1, n_{div}]$, [14]. This produced rapid complete coverage of the entire control space in a coarse distribution followed by progressive increases in density, continued until the precision limit, $2^{n_{div}}$.

The second comparison algorithm was a purely random search algorithm, which repeatedly selected control configurations at random. While random searching did not always identify the global minimum within a set number of search cycles as grid refinement did, Knepper et. all demonstrated that random search patterns could perform equal to or better in terms of time performance, [9].

## 3    Systematic Testing

The first set of tests was conducted using a contrived heuristic function to investigate the relative performance of RRHC in response to three control space parameters. Specifically, i) the number of dimensions in the control space (dimensionality), $n_{dim}$, ii) the number of grid divisions along each dimension (precision), $2^{n_{div}}$, and iii) the percentage of control space sloped towards suitable control configurations (complexity).

Here, the simulation and heuristic evaluation within each search cycle was replaced by a single calculation (Eq. 1) which produced a multidimensional quadratic bowl with a superimposed wave function (Fig. 2). The bowl guaranteed that only one global minimum would exist, centered at $\vec{d}$ (randomly selected within the control space), while the cosine wave produced a set number of local minima (complexity) according to a frequency parameter, $f$. Furthermore, the bowl slope is controlled such that the global minimum exhibited zero cost and two locally minimal solutions will existed along each dimension with cost = 0.05.

$$h(\vec{c}) = \sum_{j=1}^{n_{dim}} \left( 0.05f^2 \left( d_j - c_j \right)^2 - \cos \left( 2\pi f \left( d_j - c_j \right) \right) + 1 \right) \tag{1}$$

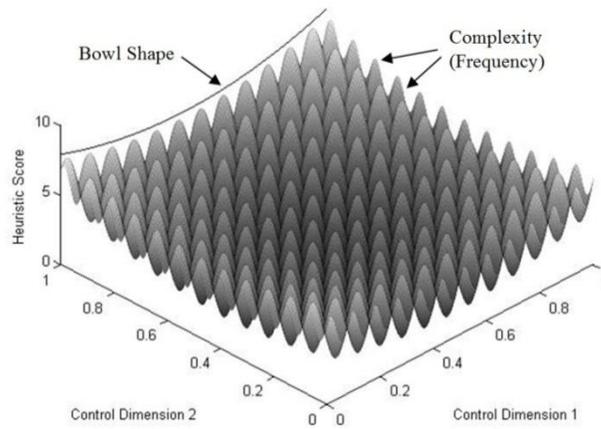

**Fig. 2.** Systematic Control Space Example (from Eq. 1 with $n_{dim} = 2$, $n_{div} = 2^8$, $f = 12$)

For each search algorithm, the test recorded the number of search cycles required to identify a control configuration with a cost below each benchmark. This process was repeated 100 times for each test/algorithm combination and the search cycle counts were averaged. These averaged results are presented below, normalized to the number of possible control configurations, $n_{config}$, per Eq. 2.

$$n_{config} = \left( 2^{n_{div}} + 1 \right)^{n_{dim}} \tag{2}$$

### 3.1 Dimensionality and Precision

The results of the dimensionality (Fig. 3) and precision (Fig. 4) tests indicate that RRHC has comparable performance at low dimensions/divisions. As the number of dimensions or divisions increases however, relative performance improves accordingly. This indicates that RRHC is most suitable for application on control problems with greater than 2 dimensions and $2^5$ divisions where it yielded significantly improved performance over the comparison algorithms (up to 1000 faster). Also of note, the performance difference was less marked with the 0.05 benchmark than the 0.0 benchmark. This is a result of the increasing percentage of the control space with costs below the benchmark value.

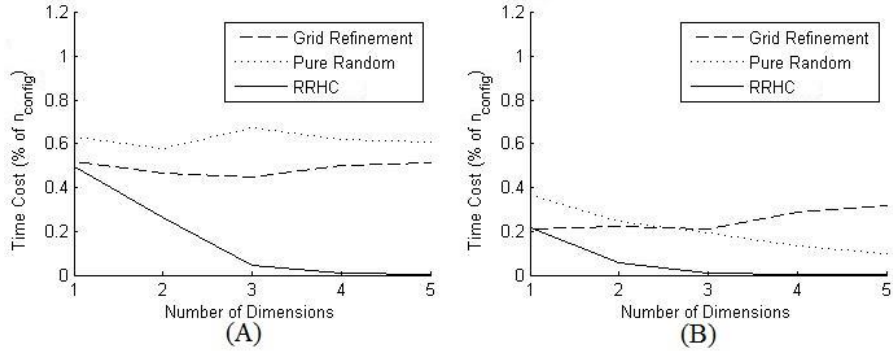

**Fig. 3.** Dimensionality Results ($n_{div} = 6$, $f = 6$): A) 0.0 Benchmark, B) 0.05 Benchmark

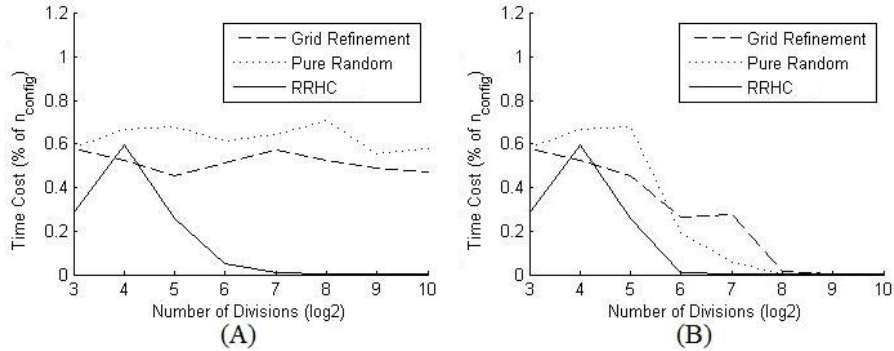

**Fig. 4.** Precision Results ($n_{dim} = 3$, $f = 6$): A) 0.0 Benchmark, B) 0.05 Benchmark

## 3.2 Complexity

Results from the complexity test (Fig. 5) demonstrate the weakness of RRHC; specifically that the algorithm performance will decrease when used with a highly stochastic or oscillatory heuristic function, due to local minima entrapment. In comparison to the other search algorithms however, hill climbing still exhibited significantly better performance, though this improvement is inversely related to complexity.

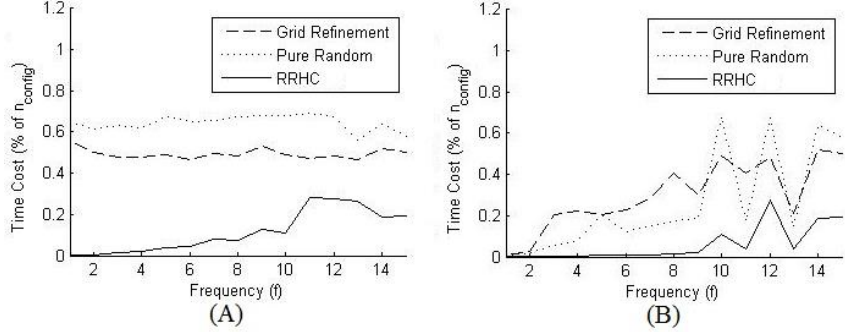

**Fig. 5.** Complexity Results ($n_{dim} = 3$, $n_{div} = 6$): A) 0.0 Benchmark, B) 0.05 Benchmark

## 4   Real-World Verification

While systematic testing provided insight into RRHC's performance, it produced very impressive results. To determine whether this performance increase would carry over into a real-world scenario, subsections of the experiments from Knepper et. al. were recreated, [9]. In these, a differential drive robot was presented with a single path environment that it must navigate. Thus, the robot needed to select a collision free set of control outputs which would advance it along the path. The control output was a set of 4 steering rates ranging from $\pm 2.1 \, rad/m$, employed in sequence for $0.3 \, m$ each. This produced a 4 dimensional control space ($n_{dim} = 4$) with steering angle rates divided into either 8 divisions ($n_{div} = 3$) for coarse control or 32 divisions ($n_{div} = 5$) for fine control.

The environment (Fig. 6) contained only the path walls, generated from perlin noise to create a curving path with multiple constrictions whose curvature was bounded to the turning rate of the robot. The average length and width of the constrictions was manipulated to produce two environments i) wide paths (Fig. 6-A, $0.05 \, m$ length and $0.05 \, m$ width avg.) and ii) thin paths (Fig. 6-B, $0.15 \, m$ length and $0.01 \, m$ width avg.).

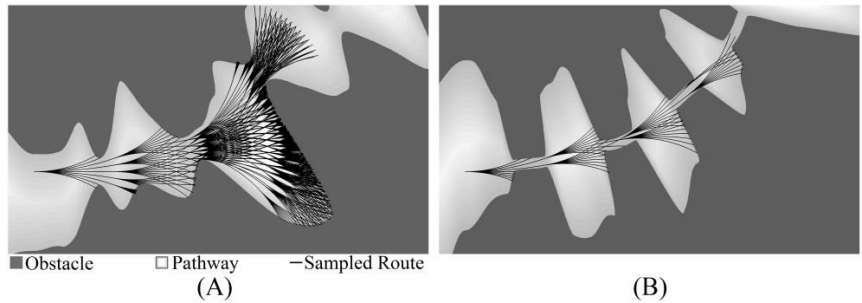

■ Obstacle   □ Pathway   — Sampled Route
(A)                 (B)

**Fig. 6.** Real-World Navigation Scenario Examples. A) Wide Paths, B) Thin Paths

As the goal of the system was to produce collision free forward motion along the path, Eq. 3 was employed as the heuristic evaluation function. It was evaluated continuously along the simulation predicted line of travel, and the minimum value of the function along that line of travel was returned to the search algorithm. At each point, the function rewarded distance travelled, $d_{travel}$, and penalized the minimum thus far observed distance between the robot and an obstacle, $d_{obs,min}$. For mathematical reasons, $d_{obs,min}$ was limited to a minimum of 0.0001.

$$h(\vec{c}) = (0.3 n_{dim} - d_{travel}) + 0.1 \left(\frac{0.01}{d_{obs,min}}\right)^2 \tag{3}$$

### 4.1 Performance on Wide Paths

Fig. 7 shows the cost, averaged over 100 trials, as it decreases over the first 5000 search cycles. For coarse control, RRHC reached global minimum first (468 cycles avg.), significantly faster than the comparison algorithms (3219 cycles for grid and 2735 cycles for random). Fine control showed similar relative performance with the hill climbing, grid refinement, and random algorithms taking an average $1.1e^5$, $5.5e^5$, and $5.0e^5$ cycles respectively. The average minimum cost identified by the fine control was 0.0285, slightly lower than that for coarse control at 0.0410. This demonstrates that fine control can yield better results given sufficient cycles.

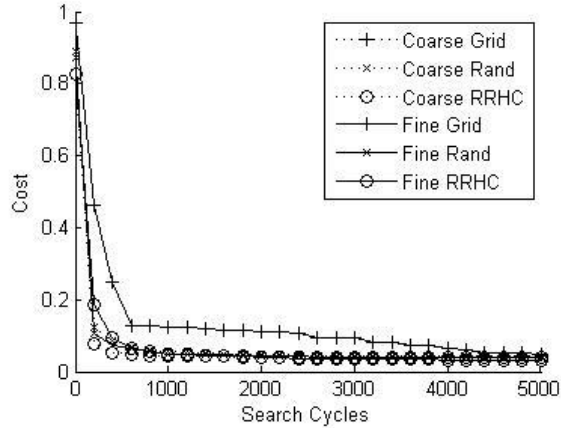

**Fig. 7.** Search Performance on Wide Paths

Surprisingly, fine RRHC began to surpass coarse RRHC after on a few cycles (approx. 1700 cycles). This indicates that although the fine control provides $1.1e^6$ possible control configurations (as compared to 6561 for coarse control) it can still be processed with comparable efficiency.

### 4.2    Performance on Thin Paths

The thin pathway performance (Fig. 8) demonstrated a significant difference between fine control (0.4527 average minimum cost) and coarse control (0.8522 average minimum cost). This disparity also caused fine RRHC to surpass coarse RRHC after only 100 cycles on average. Using fine and coarse control, RRHC required an average $4.6e^4$ and 175 cycles respectively to reach the average global minimum. Comparatively, grid refinement required an average $6.0e^5$ and 2526 cycles and pure random $4.7e^5$ and 646 cycles.

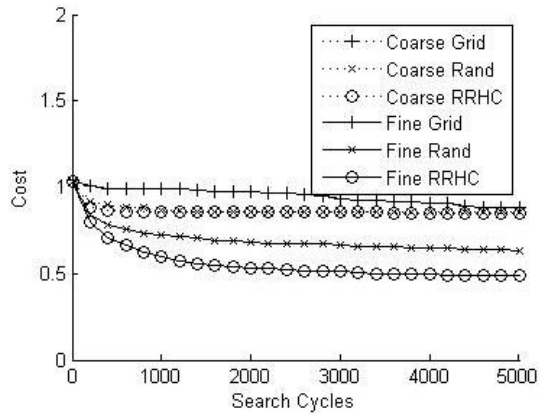

**Fig. 8.** Search Performance on Thin Paths

Taken together with the performance on wide paths, these results demonstrated a significant performance gain when employing RRHC, correlating with the systematic tests earlier. The performance gains within the real-world scenario (5 to 12 times faster) were not as dramatic as those from systematic testing (2 to 1000 times faster) however improvement by a factor of 5 remains a significant gain.

## 5    Conclusions and Future Work

The tests presented here indicate that RRHC can improve the performance of MPC. Systematic testing indicated that as the number of control dimensions and the level of control precision in each of those dimensions increases, RRHC began to significantly outperform the alternative algorithms. These performance increases carried over to a real-world test scenario, where MPC was successfully used by a simulated differential drive robot to move through a constricted path.

There remains significant investigation to be done on both MPC and the application of hill-climbing search algorithms therein. Most notably that RRHC is only one such algorithm and may not be the most suitable. In the future we intend to compare a number of hill-climbing algorithms and to conduct a wider range of real-world tests, both simulated and experimental. In the immediate future however, the application of hill climbing methods significantly improves the viability of MPC for autonomous vehicle control.